\documentclass{moriond}

\def\be{\begin{equation}}
\def\ee{\end{equation}}
\def\bea{\begin{eqnarray}}
\def\eea{\end{eqnarray}}

\newcommand{\Photo}{\includegraphics[height=35mm]{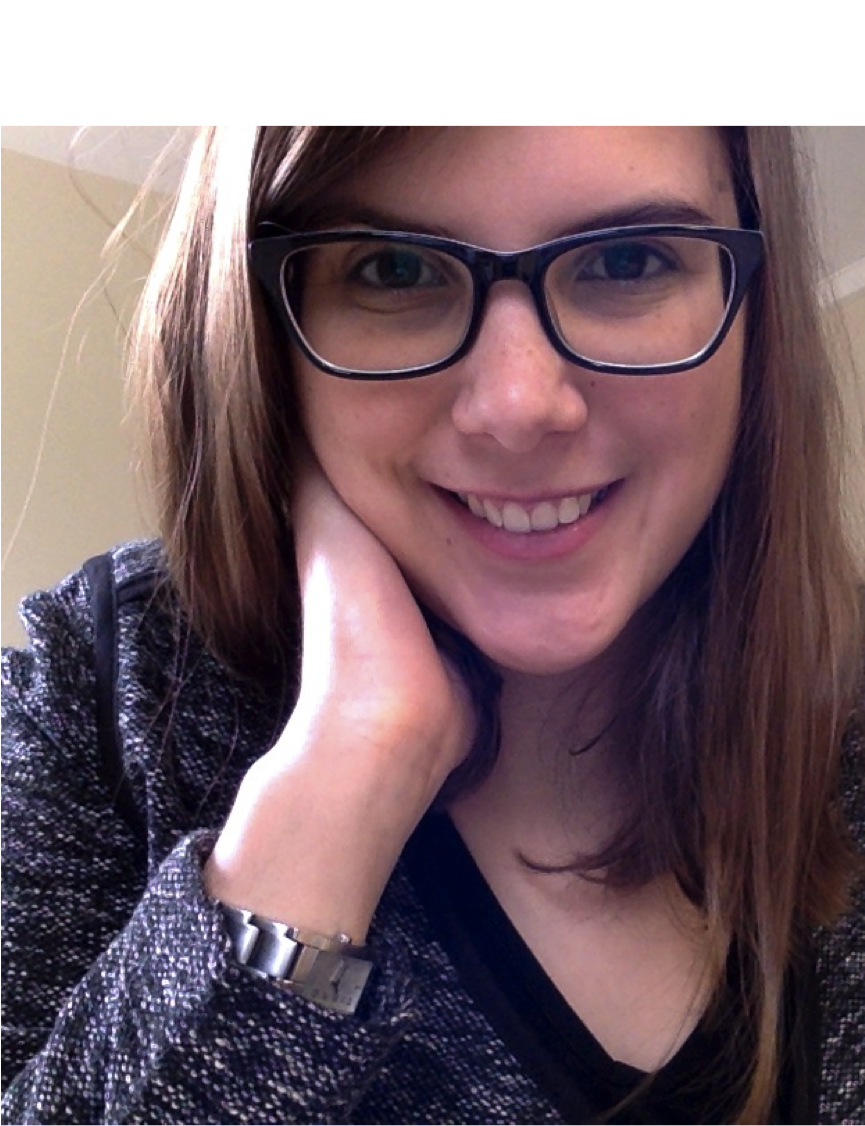}}

\begin{document}
\vspace*{4cm}
\title{ON THE STUDY OF SOLAR FLARES WITH NEUTRINO OBSERVATORIES.}

\author{ G. de Wasseige for the IceCube Collaboration }

\address{Vrije Universiteit Brussel
\\ Inter university Institute for High Energies IIHE(ULB-VUB)
\\Pleinlaan 2, 1050 Ixelles, Brussels, Belgium}

\maketitle\abstracts{
Since the end of the eighties, in response to a reported increase of the total neutrino flux in the Homestake experiment in coincidence with solar flares, neutrino detectors have searched for signals of neutrinos associated with solar flare activity.  Protons which are accelerated by the magnetic structures of such flares may collide with the solar atmosphere, producing mesons which subsequently decay, resulting in neutrinos at $\mathcal{O}$(MeV-GeV) energies.  The study of such neutrinos would  provide a new window on the underlying physics of the acceleration process.
The sensitivity to solar flares of the IceCube Neutrino Observatory, located at the geographical South Pole, is currently under study. We introduce a new approach for a time profile analysis. This is based on a stacking method of selected solar flares which are likely to be connected with pion production. An initial approach towards a neutrino search using the current IceCube experiment as well as first efforts to improve the detection efficiency in the future are presented.}

\section{Introduction}
 In 1988, the Homestake experiment observed an increase in the total number of neutrino events in possible correlation with energetic solar flares~\cite{homestake}. Bahcall predicted that if this increase were indeed due to solar flares,
this would lead to large characteristic signals in neutrino detection experiments~\cite{bahcall}. After Bahcall's prediction, neutrino detectors such as Kamiokande~\cite{kamiokande} and SNO~\cite{sno} performed several studies. Even though these detectors used different solar flare samples and analyses, they were not able to confirm
the possible signal seen by Homestake. 

Solar flares convert magnetic energy into plasma heating and the kinetic energy of charged particles such as protons. Protons injected  downwards from the coronal acceleration region will interact with the dense plasma in the lower solar atmosphere as indicated in Eq.~(1).
\begin{equation}
p\, +\,  p \,{~\rm or~} \, p\, + \, \alpha
\longrightarrow 
\left\{
\begin{array}{l}
 \pi^+ + X; \\
  \pi^0 \,+ X; \\
    \pi^- + X; \\
    \end{array}
\right.\\
\label{reaction}
\\
\begin{array}{l}
\pi^+ \longrightarrow \mu^+ + \nu_{\mu} ;~ \mu^+ \longrightarrow e^+ + \nu_e + \bar\nu_{\mu} \\
\pi^0 \longrightarrow 2 \gamma \\
\pi^- \longrightarrow \mu^- + \bar{\nu}_{\mu}; ~\mu^- \longrightarrow e^- + \bar{\nu}_e + \nu_{\mu} \\
\end{array}
\end{equation}
In addition to producing neutrinos, solar flares emit radiation across the entire electromagnetic spectrum~\cite{hudson}.
Gamma-rays are produced by both neutral pion-decay and 
secondary electron bremsstrahlung \cite{vilmer}. The neutrinos, tied to the commonly detected gamma-rays from neutral pion decay as shown in Eq.~(\ref{reaction}),
would constitute a new multi-messenger approach to study hadron acceleration in solar
flares, providing new constraints on the proton spectral index and the composition of
the accelerated flux.
Also, as pointed out by R.J. Murphy et al.~\cite{interest} concerning gamma-rays and neutrons, solar flare neutrinos would offer the potential to learn about the structure and evolution of the flare environment.
\section{Evaluation of the solar flare neutrino flux}
In order to evaluate the neutrino flux produced by a single solar flare, we have designed a Geant4~\cite{geant4} simulation of proton-nucleus interactions in the Sun's chromosphere. For this we have used the `Model of Chromospheric Flare Regions'~\cite{machado} to define the density profile of the interaction region 
and injected a proton beam in the direction of Earth hitting the chromosphere at the edge, which we indicate as \textit{tangent injection}.

\begin{figure}
	\centering
	\includegraphics[width=0.66\textwidth]{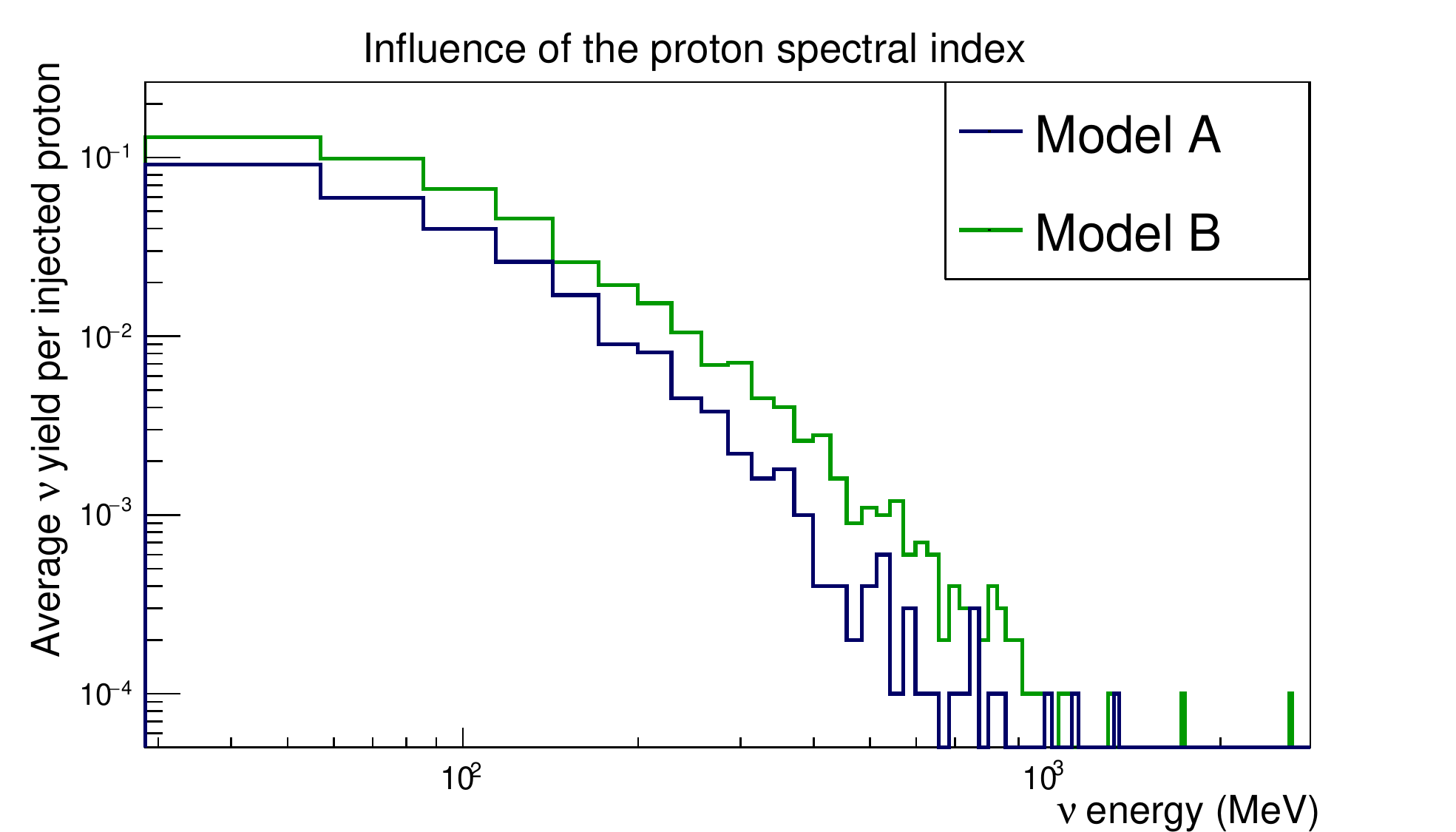}
	\caption{Influence of the spectral index of the accelerated proton flux on the solar flare neutrino yield. \textit{Model A} and \textit{Model B} are described in the text. \label{spectralindex}}
 \end{figure}

In order to be as model-independent as possible, one of the proton spectra injected in this simulation has been derived from gamma-ray observations of the June 3, 1982 event by R.J.~Murphy and R.~Ramaty~\cite{modelA}. We will call this $E^{-3.1}$ proton spectrum \textit{Model A}.
A second $E^{-2}$ proton spectrum, expected from generic Fermi shock acceleration, has also been simulated and will be referred to as \textit{Model B}.
The details of these two different models are presented in Table \ref{table-models} and their influence on the solar flare neutrino flux in direction of Earth is shown in Figure~\ref{spectralindex}.  Here it can be seen that, while both models lead to neutrino energies above 1\,MeV, the $E^{-2}$ spectrum yields a significant increase compared to \textit{Model A}.

\begin{table}[h!]
\begin{center}
\caption{\label{table-models} Details of the two models of different simulated proton spectra. The last column represents the total number of protons which have been accelerated to an energy E $>$ 30\,MeV by the solar flare. These numbers have been obtained by assuming that the total solar flare energy is the same in both models~\protect\footnotemark[1]. \setlength{\belowcaptionskip}{5pt}}
\begin{tabular} {c| c | c} \hline
Model & Proton spectrum & Total number of protons \\ \hline
Model A & $E^{-3.1}$ &$2.2\times10^{33}$ \\ 
Model B & $E^{-2}$ &$7.9\times10^{32}$
\end{tabular}
\end{center}

\end{table}

\footnotetext[1]{Even though this approximation  may not be correct for a general flare, it becomes reasonable for the solar flare sample which will be considered in this context.} 


In our studies, a pure accelerated proton flux has been assumed~\footnote[2]{The composition of the simulated atmosphere is nevertheless a mixing of Hydrogen and Helium in a ratio He:H = 1:9 as is commonly used.}. A mixed accelerated flux would increase the neutrino flux directed to Earth.

In order to evaluate the feasibility that IceCube would be able to see a signal from solar flares, we have investigated a tangent injection of a proton spectrum based on \textit{Model A} regarded here as an optimistic but realistic case. The solar flare neutrino flux is calculated for the two extreme upper cutoff values of the proton spectrum, 1\,GeV and 5\,GeV,  as presented in Table~\ref{table-flux}. The evidence for such cutoff behavior has been demonstrated by Heristchi et al.~\cite{trottet}. 
Table~\ref{table-flux} divides and presents the spectrum in two energy ranges, as relevant to detectors of different sizes: with kiloton-scale neutrino detectors sensitive to below 100\,MeV, and megaton-scale detectors sensitive to neutrinos above 100\,MeV by means of advanced data analysis techniques. 

\begin{table}[t]
\begin{center}
\caption{\label{table-flux} Expected neutrino fluence at Earth and spectrum for one single solar flare. The numbers presented in this table have been obtained from the simulation of a tangent injection of the proton spectrum described by the \textit{Model A} with an upper cutoff at 1\,GeV (a) and 5\,GeV (b).\label{table-events} \setlength{\belowcaptionskip}{5pt}}
\begin{tabular}{c |  c | c}
  \hline
Neutrino energy range &Neutrino fluence at Earth &Corresponding neutrino spectrum\\
  \hline
10 - 100\,MeV & 398$^a$ - 770$^b$ $\nu$ cm$^{-2}$& E$^0$ \\
$> 100$\,MeV & 221$^a$ - 783$^b$ $\nu$ cm$^{-2}$ & E$^{-2.3}$ \\
  \hline
\end{tabular} 
\end{center}

\end{table}

\section{IceCube and solar flare neutrinos: A novel idea for a stacking analysis} \label{icecube}\label{novel idea}
IceCube~\cite{icecube}, the neutrino observatory buried in the Antarctic ice, is made of 86 vertical strings most of which are spaced apart by 125 meters. Each string contains 60 digital optical
modules placed between 1450 and 2450 meters below the surface of ice. These modules detect Cherenkov light emitted by charged particles produced in neutrino interactions with the nearby ice or underlying bedrock. Even though IceCube has been optimized to detect TeV-scale neutrinos, it contains a denser sub-detector for studying neutrinos with lower energies called DeepCore. It consists of 15 strings and has an energy threshold of roughly 10\,GeV. The higher sensitivity of DeepCore to atmospheric neutrino oscillations and dark matter searches is achieved by placing higher-quantum-efficiency DOMs closer to each other on strings (with 7\,m spacing) and the strings themselves closer to each other (only 70\,m apart)~\cite{deepcore}.

\begin{figure}[th!]
    \centering
    \includegraphics[width=0.55\textwidth]{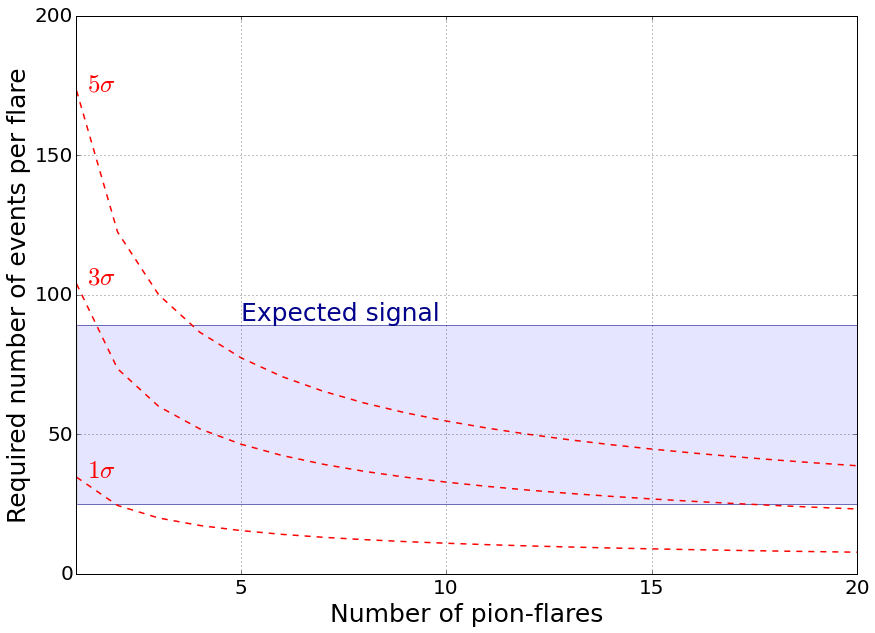}
    \caption{Comparison of the expected detected events and the required number of events per flare for a 1, 3 or 5\,$\sigma$ detection in DeepCore. The expected signal bounds have been obtained assuming the two upper cutoff values considered in Table~\protect\ref{table-events}.  A 5~$\sigma$ detection can be achieved with a sample of as few as five pion-flares in the case of \textit{Model A}, using a tangent injection and an upper cutoff of 5\,GeV.}
\label{bkgvssignal}
    
\end{figure}

The sensitivity of IceCube to solar flare neutrinos has been estimated in \cite{icrcproceeding} where it is shown that using DeepCore appears to be a promising way to detect solar flare neutrinos with energies above 100\,MeV by searching for  causally connected signal clusters within a predefined solar flare time window. 
However, even if a detection during an individual flare might be possible, a stacking of solar flare events will be more promising.
In order to maximize the detection probability, we define new criteria on the solar flare selection, rather than considering systematically every solar flare with a total number of registered photon
counts $I >I_{\small{\textrm{threshold}}}$~\cite{sno}.
The sample is restricted to solar flares for which pion-decay gamma-rays have been observed (e.g., by Fermi-LAT~\cite{pionflarefermi}). We call these specific solar flares `pion-flares'.
We use satellite data as a reference for a time-profile analysis as outlined in~\cite{nick}. For this, we intend to use the observed light curves of Fermi-LAT as a template to define a time window for neutrino observation~\cite{arxivpaper}. Using this technique, it follows that, as shown in Figure~\ref{bkgvssignal}, a few solar flares would already lead to a detection in the DeepCore region of IceCube assuming \textit{Model A}.
Considering that an estimated minimum of five pion-flares is required for a $5\,\sigma$ detection, the current list presented by the Fermi collaboration~\cite{pionflarefermi} constitutes an excellent starting point to search for neutrino emission from solar flares.


\section{Summary}

 We have designed a Geant4 simulation that considers the current parameters related to the hadron acceleration in solar flares and evaluates the solar flare neutrino flux directed to Earth.  It appears that a signal from a single flare might be seen by detecting neutrinos above 100\,MeV. However, in view of maximizing the detection probability, we have developed a new way to search for solar flare neutrinos by defining a specific solar flare sample as well as a narrow time-window in which we expect the neutrino production to occur. According to the current simulation,  five  `pion' solar flares in a stacked analysis would be required for a 5\,$\sigma$ detection in the most optimistic case for a realistic proton spectrum.
\section*{Acknowledgments}
The author wishes to thank the Inter University Attraction Pole network `fundamental interactions' for making this research possible.

\end{document}